# Sulfate phantoms to mimic NIR photoacoustic response of whole blood at selected oxygen saturations

Lea Davenet,[1,†] Arthur Billon,[1,†] Jacques Battaglia,[1] S. Lori Bridal,[1] and Jerome Gateau[1,*]

[1]*Sorbonne Université, CNRS, Inserm, Laboratoire d'Imagerie Biomédicale, LIB, F-75006, Paris, France*
[†]*These authors contributed equally.*
*\*jerome.gateau@cnrs.fr*

**Abstract:** This paper presents a practical guide to prepare inorganic solutions that mimic the photoacoustic properties of whole blood at specific levels of oxygenation in the range of 700 nm to 850 nm, both in terms of optical absorption properties and the Grüneisen coefficient. The goal is to provide aqueous solutions that can be injected in tissue-mimicking imaging phantoms to test the sensitivity and spectral capabilities of photoacoustic imaging systems operating at multiple wavelengths to estimate blood oxygenation.

## 1. INTRODUCTION

Blood, more precisely hemoglobin, is the main near-infrared (NIR) optical absorber in biological tissue [1]. Thereby, it is the primary endogenous contrast agent for photoacoustic imaging at penetration depths beyond the epithelium. Blood is also essential to biological functions because hemoglobin transports and delivers oxygen to cells. Hemoglobin's dioxygen binding capacity is central to its role in oxygen transport. It primarily occurs in two forms: oxyhemoglobin which is saturated with oxygen molecules, and deoxyhemoglobin which is desaturated. These two forms have different optical absorption spectra in the NIR range. Thereby, they can be spectrally separated and their relative concentration can be evaluated with a photoacoustic imaging system operating at multiple wavelengths using an unmixing algorithm [2]. Local evaluation of blood oxygen saturation provides functional information on oxygen availability and consumption, which can be crucial for diagnostic or therapeutic applications. A robust evaluation of blood oxygen saturation is, therefore, one of the primary objectives (and challenges) for multispectral photoacoustic imaging devices, in particular those used for functional clinical studies [3–6].

Instrument quality assessment, control and characterization must be performed on test objects with well-defined properties such as tissue-mimicking phantoms. This work presents a practical methodology for preparing aqueous solutions mimicking whole blood at different oxygen saturation levels that can be incorporated in tissue-mimicking phantoms to fill vessel-like structures of various geometries. We aim to mimic both the optical absorption coefficient and the Grüneisen coefficient of whole blood, since the product of these two coefficients is essential to accurately model photoacoustic signal amplitude.

As highlighted by the International Photoacoustic Standardization Consortium [7], photoacoustic imaging phantoms should be comprised of « chemical constituents widely available from commercial chemical vendors internationally » and have « a long-term temporal stability ». Fonseca et al. [8] previously proposed copper sulfate and nickel sulfate (in aqueous solutions) as miscible surrogate chromophores to mimic oxy- and deoxyhemoglobin in the NIR. These compounds are indeed widely available and can be purchased with a high purity grade from commercial chemical vendors. Fonseca et al. [8] verified the stability both in terms of long-term photostability under exposure with high peak power pulses and the absence of transient photobleaching. Moreover, they verified that the optical absorption coefficient for each of these compounds varies linearly with concentration and that the Beer-Lambert law is satisfied for mixtures of the two compounds. However, they did not define the relative concentrations needed to mimic the photoacoustic properties of whole blood at different oxygen saturation levels, and the spectral response of each compound did not correspond well to that of blood in the 700 – 850 nm range. Additionally, they reported the Grüneisen coefficients for different concentrations of each species, but not for mixtures.

A detailed description of the photoacoustic model of whole blood is provided, including the equations used to determine the concentrations of copper sulfate and nickel sulfate that best mimic photoacoustic response as a function of the percent oxygen saturation. The necessary parameters are then experimentally estimated through photoacoustic characterization of the mixtures. Finally, the concentrations required to obtain aqueous solutions mimicking oxygen saturation between 40% and 98% are summarized and the spectral response calculated using the photoacoustic model for whole blood is superimposed with photoacoustic spectra measured experimentally from mixtures of the sulfate salts at relative concentrations selected to mimic different levels of blood oxygen saturation.

## 2. MATERIALS AND METHODS

The first NIR window in biological tissue (optical wavelength range from 650 nm to 1000 nm) corresponds to the typical operational range of commercial photoacoustic imaging scanners that are currently available and able to image tissues at centimeter-scale depths [6]. In the imaging mode referred to as photoacoustic tomography, a reconstructed photoacoustic image maps the spatial distribution of the initial acoustic pressure rise based on the reception of ultrasound waves generated due to optical absorption. In the thermal and stress confinement regimes, usually satisfied in photoacoustic imaging systems within this wavelength range using nanosecond pulsed lasers, this initial local pressure increase $p_0$ can be expressed at each optical excitation wavelength, λ, by eq. 1 [9].

$$p_0(\lambda) = \phi(\lambda) \cdot \Gamma_{water} \cdot \theta^{PA}(\lambda) \qquad (1)$$

where $\phi(\lambda)$ is the local optical flux density and $\theta^{PA}(\lambda)$ is the photoacoustic coefficient further defined in eqs. 2 and 3. For the sake of readability, the spatial dependence of $p_0$, $\phi$ and $\theta^{PA}$ was not indicated but the quantities are defined locally and have spatial distributions in the illuminated volume. $\Gamma_{water}$ is the Grüneisen coefficient of water, which is independent of the optical wavelength but depends on the equilibrium temperature. Here, water serves as a reference because it is the primary component of blood plasma.

The photoacoustic coefficient depends on the local properties of the optically absorbing medium and can be defined as [9]:

$$\theta^{PA}(\lambda) = \eta_{PA}(\lambda) \cdot \mu_a(\lambda) \qquad (2)$$

where $\mu_a(\lambda)$ is the local optical absorption coefficient and $\eta_{PA}(\lambda)$ is the photoacoustic generation efficiency (PGE) which can further be expressed as in eq. 3:

$$\eta_{PA}(\lambda) = E_{pt}(\lambda) \cdot \frac{\Gamma}{\Gamma_{water}} \qquad (3)$$

where $E_{pt}$ is the photothermal conversion efficiency coefficient characterizing light conversion to heat and $\Gamma$ is the local Grüneisen coefficient of the medium that characterizes the conversion efficiency of heat energy into the initial pressure rise.

From this set of equations (eqs. 1-3), we can derive that the photoacoustic parameters that need to be well mimicked by the whole blood mimicking solutions are $\mu_a$, $E_{pt}$ and $\Gamma$, or at least the product of those three parameters. Optical absorption of whole blood in the NIR can be assumed to be entirely due to hemoglobin molecules [10], which are not fluorescent and, therefore, have a photothermal conversion efficiency coefficient of 100% so that $E_{pt}=1$. Then, the PGE of whole blood is independent of the optical wavelength in the NIR and is equal to the Grüneisen coefficient of whole blood relative to water.

$$\eta_{PA}^{wb} = \frac{\Gamma_{wb}}{\Gamma_{water}} \qquad (4)$$

In the following sub-sections, we first review the current state of knowledge related to the optical absorption coefficient and the Grüneisen coefficient of whole blood at different levels of blood oxygen saturation. Then, we determine the optical wavelengths that are most commonly-used to measure blood oxygenation in photoacoustic imaging and the corresponding wavelength range. Finally, we explain how we mimic the photoacoustic properties of whole blood in this wavelength range using mixtures of copper sulfate and nickel sulfate in solution.

## 2.1 Oxygenation and photoacoustic properties of whole blood

The optical absorption coefficient of whole blood depends on its oxygen saturation. Additionally, both the absorption coefficient and the Grüneisen coefficient depend on the hematocrit (volume percentage of red blood cells). We review here the different parameters considered (hematocrit, oxygen saturation levels and Grüneisen coefficient) and choices that we made to model the photoacoustic properties of whole blood.

### 2.1.1 Blood oxygen saturation levels

The spectral variations of blood optical absorption depend on oxygen saturation. Evaluating the oxygen saturation of blood consists in determining the fraction of oxyhemoglobin relative to the sum of oxy and deoxy-hemoglobin. The oxygen saturation is noted as $SO_2$ can be expressed as:

$$SO_2 = \frac{c_{HbO_2}}{c_{Hb} + c_{HbO_2}} \qquad (5)$$

where $c_{HbO_2}$ is the molar concentration of oxyhemoglobin (notated $HbO_2$) and $c_{Hb}$ is the molar concentration of deoxyhemoglobin (notated Hb). In the following, $SO_2$ is expressed as a percentage in the text in accordance with common practice in scientific literature. However, it is used in its fractional form (comprised between 0 and 1) when it is a computing variable in eqs 6 and 12.

The arterial oxygen saturation $SaO_2$ refers to $SO_2$ in arteries. Most often, it is estimated indirectly by non-invasive optical measurements through pulse oximetry and values for healthy individuals are in the range of 95% to 100% [11]. This type of optical measurement based on light absorption advantageously uses the fact that $HbO_2$ and Hb have different light absorption spectra. Pulse oximetry typically evaluates absorption at two optical wavelengths (660 nm and 940 nm), but has limitations that are mainly due to the potential presence of dyshemoglobin which is incapable of reversible association with oxygen [12]. For instance, in a cohort of 631 patients undergoing cardiopulmonary exercise testing, the average $SaO_2$ was determined to be 94% ± 2% using invasive arterial blood gas analysis while it was evaluated to be significantly higher (98% ± 2%) by pulse oximetry [13]. Mixed venous oxygen saturation $SvO_2$ refers to the oxygen content of the blood returning to the heart after oxygen delivery to tissues. Normal $SvO_2$ values are in the range of 70 to 80%. Continuous measurement of $SvO_2$ can be obtained using a fiber-optic catheter inserted in the pulmonary artery and optical reflection spectrophotometry [11]. Moreover, local venous oxygen saturation depends on the oxygen extraction of the organ and patient conditions [3]. To cover a broad range of possible conditions, we chose to mimic whole blood with $SO_2$ % between 40% and 98%.

### 2.1.2 Optical absorption of whole blood

The optical absorption of whole blood is mainly due to absorption by oxygenated and deoxygenated hemoglobin molecules. To model the absorption coefficient as a function of the SO$_2$ level, we used the spectra compiled by Bosschaart et al [10] for SO$_2$= 0%, $\mu_a^{wb}(0\%, \lambda)$, and for SO$_2$ > 98 %, $\mu_a^{wb}(98\%,)$. These spectra correspond to a hematocrit of 45% and are presented in Fig.1 (c). Assuming that the spectra for SO$_2$ > 98 % corresponds to an SO$_2$ equal to 98% and using the Beer-Lambert law, we can calculate the absorption coefficient of whole blood $\mu_a^{wb}$ at any given SO$_2$ and wavelength, $\lambda$, using the set of eqs. 6-7.

$$\mu_a^{wb}(SO_2, \lambda) = (1-SO_2) \cdot \mu_a^{wb}(0\%, \lambda) + SO_2 \cdot \mu_a^{wb}(100\%, \lambda) \quad (6)$$

$$\mu_a^{wb}(98\%, \lambda) = 0.98 \cdot \mu_a^{wb}(100\%, \lambda) + 0.02 \cdot \mu_a^{wb}(0\%, \lambda) \quad (7)$$

### 2.1.3 The Grüneisen coefficient of whole blood

Several studies have reported estimated values of the Grüneisen coefficient for whole blood. Yao *et al.* performed measurements at 22°C using diluted bovine red blood cell suspensions [15]. Based on their results, for a hematocrit of 45%, which corresponds to a hemoglobin mass concentration of 150 g/L, we can derive a value of $\Gamma_{wb} \approx 0.174$. Villanueva-Palero *et al.* reported measurements performed on whole human blood and obtained $\Gamma_{wb} \approx 0.167$ at 22°C [16]. No evidence of a specific dependency of the Grüneisen coefficient with the oxygen saturation was reported [16,17]. Additional estimations at 37°C led to $\Gamma_{wb} \approx 0.226$. However, this is the only measurement found in the literature that was performed at 37°C.

Photoacoustic phantoms are usually used at room temperature, either with whole blood or with surrogate solutions. We, therefore, chose to set the modeled Grüneisen coefficient of whole blood to $\Gamma_{wb} \approx 0.17$ in coherence with results published in [15] and [16]. For comparison, the Grüneisen coefficient of plasma at 22°C is $\Gamma_{plasma} \approx 0.12$ [16]. The value of 0.12 is also a good estimation of the Grüneisen coefficient of water at 22°C. This similarity confirms the relevance of using water as a reference in eq. 1.

### 2.1.4 The photoacoustic generation efficiency and photoacoustic coefficient of whole blood as a function of oxygen saturation

The photoacoustic generation efficiency (PGE) of whole blood was defined in eq. 4 as the ratio between the Grüneisen coefficients of whole blood and water. At 22°C, we set the PGE of whole blood equal to: $\eta_{PA}^{wb} = 0.17/0.12 \approx 1.417$

Consequently, according to equation 2, 4 and 6, the photoacoustic coefficient of whole blood at any given SO$_2$ and wavelength $\lambda$ can be modeled by:

$$\theta_{wb}^{PA}(SO_2, \lambda) = \eta_{PA}^{wb} \cdot \mu_a^{wb}(SO_2, \lambda) \quad (8)$$
$$= 1.417 \cdot \left((1-SO_2) \cdot \mu_a^{wb}(0\%, \lambda) + SO_2 \cdot \mu_a^{wb}(100\%, \lambda)\right)$$

### 2.1.5 Commonly-used optical wavelengths to determine the SO$_2$ in PA imaging

We conducted a literature review to determine the most commonly-used optical wavelengths for spectral separation of oxyhemoglobin and deoxyhemoglobin using photoacoustic imaging. We selected 23 articles published between 2013 and 2025, in which an assessment of the oxygen saturation was performed *in vivo* or *ex vivo* in preclinical and clinical studies with NIR lasers [3–5,18–37]. A total of 24 different sets of optical wavelengths were extracted. Fig. 1 (a) presents the histogram of the number of studies using each range of wavelengths computed for a bin width of 10 nm. Six different wavelengths (center of the bins) stand out from the others: 700 nm, 730 nm, 750 nm, 760 nm, 800 nm and 850 nm. The median number of wavelengths used was 5, with a minimum of 2 and a maximum of 28. The wavelengths around 750 nm and around 760 nm aim at probing the local absorption peak of deoxyhemoglobin which can be noticed on the absorption spectrum of whole blood at SO$_2$ =0% in Fig. 1(c). Because the relative variation of the absorption of whole blood between 750 nm and 760 nm is only 4% for $SO_2$ =0%, these two wavelengths do not probe a significantly different feature. Therefore, we selected the five wavelengths: $\lambda_1^{Hb}$ = 700 nm, $\lambda_2^{Hb}$ = 730 nm, $\lambda_3^{Hb}$ = 760 nm, $\lambda_4^{Hb}$ = 800 nm and $\lambda_5^{Hb}$ = 850 nm as denoted by vertical dashed lines in Fig. 1(c). From this analysis, we conclude that the most relevant wavelength range for the determination of the oxygen saturation in photoacoustic imaging is between 700 nm and 850 nm. The middle of this range is $\lambda_m = 775\ nm$.

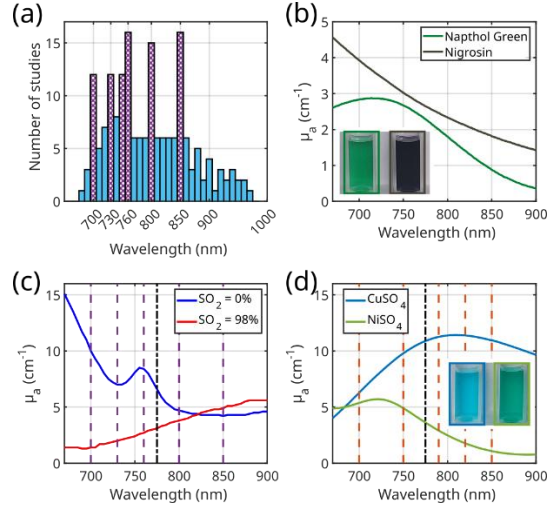

Fig. 1. (a) Histogram of the optical wavelengths used to determine the oxygen saturation in 24 published studies. The most commonly used wavelengths have been highlighted with a purple hatch pattern. (b) Optical absorption spectra of the reference solutions of Naphthol Green and nigrosin. (inset) picture of the solutions. (c) Optical absorption spectra of whole blood for $SO_2$ = 0% and 98% obtained from [10]. The vertical purple dashed lines correspond to 5 of the most commonly used optical wavelengths for the estimation of oxygen saturation. The black dashed line indicates the optical wavelength 775 nm. (d) Optical absorption spectra of the stock solutions of $CuSO_4$ and $NiSO_4$. The vertical orange dashed lines correspond to 5 of the most used optical wavelengths but mirrored with respect to 775 nm. (inset) picture of the solutions.

## 2.2 Mimicking photoacoustic properties of oxy- and deoxyhemoglobin with two sulfate salts

### 2.2.1 Optical and photoacoustic properties of copper and nickel sulfate solutions

Copper and nickel sulfate salts are water soluble. Although the compounds were used in their hydrated form ($CuSO_4 \cdot 5H_2O$ and $NiSO_4 \cdot 6H_2O$), they are referred to simply as copper sulfate, $CuSO_4$, and nickel sulfate, $NiSO_4$ for better readability. Both the solutions of $CuSO_4$ and $NiSO_4$ absorb light in the NIR, but they have different spectra. Fig. 1 (d) presents the absorption spectra for stock solutions with mass concentrations of 100 g.L$^{-1}$ (0.40 mol.L$^{-1}$) for $CuSO_4$ and 300 g.L$^{-1}$ (1.14 mol.L$^{-1}$) for $NiSO_4$, respectively. The solutions can be diluted with water and can be mixed. The optical absorption of the mixture obeys the Beer-Lambert law so that the absorption coefficient of a mixture is given by the following equation as previously shown by Fonseca *et al* [8] :

$$\mu_a^{mix}(c_{CuSO_4}, c_{NiSO_4}, \lambda) = \ln(10) \cdot (\varepsilon_{CuSO_4}(\lambda) \cdot c_{CuSO_4} + \varepsilon_{NiSO_4}(\lambda) \cdot c_{NiSO_4}) \qquad (9)$$

where $c_{CuSO_4}$ and $c_{NiSO_4}$ are the molar concentrations and $\varepsilon_{CuSO_4}$ and $\varepsilon_{NiSO_4}$ are the molar absorption coefficients of $CuSO_4 \cdot 5H_2O$ and $NiSO_4 \cdot 6H_2O$, respectively. It can already be deduced from Fig. 1 (d) that $\varepsilon_{CuSO_4}$ is much larger than $\varepsilon_{NiSO_4}$ in the NIR.

Regarding the photoacoustic generation efficiency (PGE), the solutions are photostable and the photothermal conversion efficiency is equal to $E_{pt}=1$ [8,9]. Fonseca *et al* [8] provided an empirical formula approximating the Grüneisen coefficient of a mixture in solution as a linear function of the salt concentration. This formula was further validated here for the investigated concentration range (section 3.3), and implies that the PGE can be written as:

$$\eta_{PA}^{mix}(c_{CuSO_4}, c_{NiSO_4}) = \frac{\Gamma_{mix}}{\Gamma_{water}} = 1 + \beta_{CuSO_4} \cdot c_{CuSO_4} + \beta_{NiSO_4} \cdot c_{NiSO_4} \quad (10)$$

where $\beta_{CuSO_4}$ and $\beta_{NiSO_4}$ (unit: M$^{-1}$ or L.mol$^{-1}$) are empirically determined factors (with positive values) that weights the influence of the salt concentration on the Grüneisen coefficient of the mixture. The PGE of the mixture equals one for water (salt concentration equal to zero) and increases with the increasing salt concentration. The values of these $\beta$ factors were previously determined independently on solutions of either CuSO$_4$ or NiSO$_4$ by Fonseca *et al.* [8] and also by our group [9]. We also determined that the $\beta$ factors can be temperature-dependent. However, the $\beta$ factors were not systematically tested on mixtures. Moreover, a discrepancy between the reported values of the $\beta$ factors called for the additional experiments [9] that were performed in section 3.3.

From eq. 2, the photoacoustic coefficient of the mixture in solution can be written as:

$$\theta_{mix}^{PA}(c_{CuSO_4}, c_{NiSO_4}, \lambda) = \eta_{PA}^{mix}(c_{CuSO_4}, c_{NiSO_4}) \cdot \mu_a^{mix}(c_{CuSO_4}, c_{NiSO_4}, \lambda) \quad (11)$$

And from eqs. 9-10, it can be deduced that $\theta_{mix}^{PA}$ is a quadratic function of the concentrations $c_{CuSO_4}$ and $c_{NiSO_4}$ of the two sulfate salts.

### 2.2.2 Determining the hemoglobin-mimicking analogs and their concentrations for blood-mimicking mixtures

Fonseca *et al* [8] proposed use of NiSO$_4$ as an analog of Hb and CuSO$_4$ as an analog of HbO$_2$. However, the similarity of the spectral shapes was found to be poor in the range 700 nm to 850 nm (Fig. 1(c-d)). Moreover, one can notice a similarity between the spectral shapes of HbO$_2$ and NiSO$_4$ provided a flip is made with respect to $\lambda_m$, the central wavelength of the considered range. Therefore, we propose a different approach in the wavelength range 700 nm to 850 nm with two significant modifications:
1) mirroring the optical wavelengths with respect to 775 nm when performing a multispectral photoacoustic acquisition with a mixture of CuSO$_4$ and Ni SO$_4$ compared to an acquisition with whole blood.
2) stating that NiSO$_4$ is the analog of HbO$_2$ and CuSO$_4$ is the analog of Hb.

The first modification corresponds to the following transformation between the wavelengths $\lambda^{analog}$ for the PA optical excitation of the analogs and the wavelengths $\lambda^{Hb}$ for the PA excitation of the hemoglobin: $\lambda^{analog} = 1550\ nm - \lambda^{Hb}$. Therefore, the analog PA phantom-assessment wavelengths corresponding to the spectral features of blood probed by the five most-commonly-used wavelengths currently applied for the discrimination of deoxy- and oxyhemoglobin become $\lambda_1^{analog}=850\ nm$, $\lambda_2^{analog}=820\ nm$, $\lambda_3^{analog}=790\ nm$, $\lambda_4^{analog}=750\ nm$ and $\lambda_5^{analog}=700\ nm$. These wavelengths have been marked with orange vertical lines in Fig. 1 (d). The second modification provides a link between the molar concentration of NiSO$_4$ and CuSO$_4$ and the modeled oxygen saturation SO$_2$ of whole blood according to eq. 12.

$$SO_2 \cdot c_{CuSO_4} = (1 - SO_2) \cdot c_{NiSO_4} \quad (12)$$

To be able to determine the analog concentrations, we added the following constraint:

$$\theta_{mix}^{PA}(c_{CuSO_4}, c_{NiSO_4}, \lambda_3^{analog}=790\ nm) = \theta_{wb}^{PA}(SO_2, \lambda_3^{Hb}=760\ nm) \quad (13)$$

The wavelength $\lambda_3^{Hb}$ was chosen because of its proximity to $\lambda_m$ and because almost all the selected studies (23 out of the 24) used at least one wavelength between 745 nm and 765 nm.

The system of polynomial equations given by eq. 12 and eq. 13 can be solved analytically.

$$c_{CuSO_4} = \frac{\sqrt{D}-B}{2A} \text{ and } c_{NiSO_4} = \frac{\sqrt{D}-B}{2A} \cdot \frac{SO_2}{1-SO_2} \tag{14}$$

with $B = \varepsilon_{CuSO_4}(\lambda_3^{analog}) + \varepsilon_{NiSO_4}(\lambda_3^{analog}) \cdot \frac{SO_2}{1-SO_2}$

$$A = B \cdot (\beta_{CuSO_4} + \beta_{NiSO_4} \cdot \frac{SO_2}{1-SO_2})$$

$$D = B^2 - 4 \cdot A \cdot \theta_{wb}^{PA}(SO_2, \lambda_3^{Hb}=760\ nm)/\ln(10)$$

In the results section (section 3), we present the experimentally determined quantities $\varepsilon_{CuSO_4}(\lambda_3^{analog})$ and $\varepsilon_{NiSO_4}(\lambda_3^{analog})$ as well as $\beta_{CuSO_4}$ and $\beta_{NiSO_4}$.

## 2.3 Experimental methods

### 2.3.1 Chemical products

Nickel(II) sulfate hexahydrate ($NiSO_4 \cdot 6H_2O$, 227676-500G, CAS: 10101-97-0, $M_w$ = 262.85 g.mol$^{-1}$, ACS reagent, ≥98%) and copper(II) sulfate pentahydrate ($CuSO_4 \cdot 5H_2O$, 209198-500G, CAS: 7758-99-8, $M_w$ = 249.69 g.mol$^{-1}$, ACS reagent, ≥98.0%) were purchased from Sigma-Aldrich (St. Louis, MO, USA) and were used to prepare the blood mimicking solutions.

Nigrosin (CAS: 8005-036) was obtained from Sigma-Aldrich (198285-25G, $M_w$ = 202.21 g.mol$^{-1}$) and from Thermo Fisher Scientific (Waltham, MA, USA) (189480250, $M_w$ = 616.49 g.mol$^{-1}$, high purity biological stain). Naphtol green B was purchased from Thermo Fisher Scientific (A18268.14, CAS: 19381-50-1, $M_w$ = 881.474 g.mol$^{-1}$). These dyes were used to obtain solutions with a PGE equal to one.

All solutions were prepared with ultrapure water (Resistivity > 18 MΩ.cm$^{-1}$, Purelab Option Q, ELGA LabWater).

### 2.3.2 Spectrophotometry

Optical absorbance was measured using a 0.2 cm thick quartz cuvette (QS 10.00 Hellma) with a spectrophotometer (VWR P4 Spectrophotometer, VWR, Leuven, Belgium). Blank was performed with ultrapure water. All measurements were performed at room temperature - between 20°C and 25°C.

Since all solutions were clear (no turbidity), the absorption coefficient (in cm$^{-1}$) was computed from the absorbance measurement Abs:

$$\mu_a(\lambda) = \frac{\text{Abs}(\lambda)}{0.2} \cdot \ln(10) \tag{15}$$

### 2.3.3 Stock and reference solutions

Dye solutions were prepared to yield maximum absorption coefficients between 2 and 4 cm$^{-1}$ in the range 700 nm to 850 nm. Nigrosin solutions were prepared to obtain an absorption coefficient of 4 cm$^{-1}$ at 700 nm (Fig. 1 (b)). For the Sigma-Aldrich powder, the mass concentration required to provide this value was 144 mg.L$^{-1}$. Naphtol green solution had a maximum absorption coefficient of 2.8 cm$^{-1}$ at 720 nm (Fig. 1(b)). The low dye concentrations compared to the concentrations of $CuSO_4$ and $NiSO_4$ in the mixtures and the solubility of the dyes ensure that the Grüneisen coefficient of the dye solutions equals that of water. Furthermore, the solutions are photostable and their photothermal conversion efficiency is equal to one (no fluorescence or other deexcitation pathways). Their optical absorption coefficient was determined to be independent of the temperature in the range 15°C to 35°C.

A stock solution of $CuSO_4$, used to prepare the mixtures, was obtained by pouring 2.50 g of solid crystals in a 25.0 mL volumetric flask, before gradually adjusting the volume with water. This stock solution had a mass concentration of 100 g.L$^{-1}$. In addition, another solution was prepared at 50 g.L$^{-1}$ using a 50.0 mL flask to obtain a second mass concentration from the solid crystals. The same protocol was used to obtain a stock solution of $NiSO_4$ at 300 g.L$^{-1}$ (7.50 g in 25 mL) and a solution at 150 g.L$^{-1}$.

Finally, a solution of $CuSO_4$ at 0.25 mol.L$^{-1}$ was used to calibrate the photoacoustic spectrometer together with the Naphthol green solution. These two solutions are referred to as the reference solutions for the photoacoustic spectrometer. Nigrosin solutions were introduced to provide absorbing samples without sulfate salts in the measurement series, and to serve as calibration checks for the photoacoustic spectrometer.

### 2.3.4 Mixtures

Four batches of mixtures were produced from freshly prepared stock solutions. Batch #1 comprised 8 mixtures corresponding to $SO_2$ : 0%, 40%, 50%, 60%, 70%, 80%, 90%, and 98% and was prepared using the β-factors given by Fonseca *et al* [8]: $\beta_{CuSO_4}^{preparation\ \#1}=0.71\ M^{-1}$ and $\beta_{NiSO_4}^{preparation\ \#1}=0.32\ M^{-1}$. Batches #2 to #4 comprised 10 mixtures corresponding to $SO_2$: 0%, 40%, 50%, 60%, 70%, 80%, 85%, 90%, 94% and 98%, which correspond to an extended range of physiological oxygenation saturation (see section 2.1.1). It is worth noting that each batch was prepared with a different set of β-factors based on the measurements of the previous batch and the maturity of our data processing analysis. Therefore, the concentration of the mixtures varied from one batch to another.

Concentrations of $CuSO_4$ and $NiSO_4$ needed for each modeled oxygen saturation were calculated from eq. 14 and converted into volumes to extract from stock solutions to reach a total volume of 5 mL for each mixture. The computed volumes were rounded to the nearest volume with a precision of 0.01 mL and transferred with a 1 mL pipette. The remaining volume was completed with ultrapure water

### 2.3.5 Quantitative photoacoustic spectrometry

Quantitative photoacoustic (PA) spectra of the solutions were measured with a calibrated PA spectrometer described in detail by Lucas *et al.* [9,38]. Briefly, optical excitation was generated with a tunable (680−980 nm) nanosecond Laser (pulse repetition rate: 20 Hz, pulse width 6 - 7 ns, SpitLight 600OPO, Innolas, Germany). The optical wavelength was scanned from 680 nm to 970 nm in 10 nm steps using the per-pulse tunability of the laser and for a total of 15 wavelength scans. Ultrasound (US) detection was performed with a linear US array (L7−4, ATL, USA) driven by a programmable US machine (Vantage, Verasonics, USA). Samples were injected into four PTFE tubes (inner diameter: 0.3 mm, wall thickness: 0.15 mm, S1810-04, Bola, Germany) with a 1 mL syringe (approximately 100 μL were injected per 50 cm long tube). Excitation light on the sample had a maximum fluence of 3.5 mJ·cm$^{-2}$, and 1.5 cm of the tube was illuminated. The tube and the US detector were immersed in a thermostatic water tank (T100-ST12 Optima, Grant, UK) maintained at 25 °C or 20 °C. The water bath temperature was monitored throughout the entire experiment (HI98509, Hanna instruments, Lingolsheim, France).

The same four tubes were used for successive sample injections. We verified that no detectable absorption remained in the tubes between two samples by systematically measuring background signals with the tube filled with ultrapure water. This process served a dual process: (1) these background signals were used as a blank for PA measurements, and (2) the water flush rinsed the tubes. No noticeable changes in the measured spectra were observed over the 15 wavelength scans, and the signals were then averaged to increase the signal-to-noise ratio. Each sample was injected twice in the four tubes for a total of 8 measurements. For Batch #3 and Batch #4, unless otherwise stated, a measurement series comprised a total of 16 samples: the 10 mixtures, the 2 stock solutions, the additional solutions of $CuSO_4$ and $NiSO_4$ and the 2 solutions of nigrosin (see section 2.3.3). For Batch #1 and Batch #2, the series comprised only the mixtures. The thermalization of all samples at the temperature of the water bath was performed by immersing the sample containers (15 mL conical centrifuge tube with a screw-cap) at least 30 minutes before the injection. Indeed, the Grüneisen coefficient of the solution is temperature-dependent, necessitating precise temperature control.

In addition to the tested samples, the two reference solutions (see section 2.3.3) were measured for the calibration of the PA spectrometer. Each of the reference solutions was repetitively injected in each of the four tubes during the experiment: twice before all the tested samples and twice every three to four samples. A total of 10 measurements per tube and per reference solution was collected for Batches #3 and #4. These repetitive measurements allow for the detection of potential drifts, and ensure a robust calibration and a precise estimation of the PGE [38]. One acquisition series with drifts was fully discarded. A complete series of 16 samples resulted in a total of 78 acquisitions and lasted around 5 hours.

Once the acquisition was complete, signal processing was initiated with the computation of the calibration coefficient using the measurements of the two reference solutions following the method described in ref [38]. The first reference solution was the solution of $CuSO_4$ at a concentration of 0.25 mol.L$^{-1}$ and allowed the per-wavelength calibration thanks to its strong absorption throughout the considered NIR wavelength range. The second one was the solution of naphtol green B and allowed for the global amplitude calibration through the determination of the Grüneisen coefficient of the first reference solution $\eta_{PA}^{Calibration}$. This is the first time that naphtol green B is presented as a reference solution for photoacoustic spectrometry. This solution was preferred over the previously used nigrosin solution [9,38], because it is less prone to adsorption on the tubes' inner surface which facilitates tube cleaning with a water flush. However, only the optical wavelengths between 680 nm and 820 nm were used to determine $\eta_{PA}^{Calibration}$, due to the weaker absorption coefficient (Fig. 1(b)) of naphtol green at higher wavelengths. To validate this reference solution, we had verified that the measured PGE of the nigrosin solution was equal to one when the naphtol green solution was used as a reference, which is further demonstrated in section 3.2. The uncertainty for $\eta_{PA}^{Calibration}$ was evaluated here as equal to: $\frac{\Delta \eta_{PA}^{Calibration}}{\eta_{PA}^{Calibration}}$=0.7 % using the formula given in ref [33].

For each sample, the PA spectra were computed and converted to spectroscopic units using the calibration. The median value and the median absolute deviation with a scaling factor 1.4826 were computed over the 8 measurements to obtain the photoacoustic coefficient $\theta^{PA}(\lambda)$ and its deviation. In this paper, we call MAD the median absolute deviation with a scaling factor 1.4826.

For each sample, the PGE was computed at the optical wavelengths $\lambda \in [680nm; 920\ nm]$:

$$\eta_{PA}(\lambda) = \frac{\theta^{PA}(\lambda)}{\mu_a(\lambda)} \quad (16)$$

together with its uncertainty $\frac{\Delta \eta_{PA}(\lambda)}{\eta_{PA}(\lambda)}$ [38]. The wavelength interval $[\lambda_1^{Hb}; \lambda_5^{Hb}]$ is fully included, and measurements performed for optical wavelengths above 920 nm were discarded due to degraded performance of our instrument induced by the increase in the absorption of water. Then, the median PGE value $\widetilde{\eta_{PA}}$ was obtained from the 25 evaluations of $\eta_{PA}(\lambda)$, at each different wavelength.

### 2.3.6 Evaluation of the β factors

The PGE values $\widetilde{\eta_{PA}}$ for all the samples and $\eta_{PA}^{Calibration}$ were used to determine the $\beta$ factors. The values of $c_{CuSO_4}$ and $c_{NiSO_4}$ for each mixture were determined from the dilution of the stock solutions. The Matlab(R) function *fitlm*, which returns a linear regression model fit to the data, was used to estimate the $\beta$ factors and the standard errors of the coefficients. The coefficient of determination, the root mean square error and the maximum relative error were computed to determine the goodness of fit.

## 3. Results

### 3.1 Molar absorption coefficients at 790 nm

The molar absorption coefficients at 790 nm were computed independently for $CuSO_4$ and $NiSO_4$ from solutions with only one compound. Over the different batches, the optical absorption at 790 nm was measured for at least 4 different concentrations and a total of 9 solutions of each compound. A linear regression of the optical absorption at 790 nm as a function of concentration led to: $\varepsilon_{CuSO_4}(790\,nm) = 12.05 \pm 0.04\ mol^{-1}.L.cm^{-1}$ and $\varepsilon_{NiSO_4}(790\,nm) = 1.10 \pm 0.01\ mol^{-1}.L.cm^{-1}$. These coefficients are in agreement with previously published results [8].

It is worth noting that the optical absorption spectra of Batch #2 mixtures were measured again 4 months after preparation and storage at room temperature in sealed containers (centrifuge tubes with screw caps). The spectral shape was well superimposed with the measurements performed on the freshly prepared solutions- indicating a long-term stability.

### 3.2 PA spectra and PGE of the samples

Fig. 2(a) presents a photograph of 5 mixtures (from Batch #4) side-by-side with concentration ratios corresponding to eq. 12 and values of $SO_2^{analog}$ distributed throughout the whole range. The color gradient between light blue (0%) and green (98%) is visible. Fig. 2(b-h) illustrates the optical absorption spectra of the different solutions of Batch #3, together with their photoacoustic spectra. Additionally, the absorption spectra multiplied by the computed median PGE $\widetilde{\eta_{PA}}$ is displayed. The photoacoustic coefficients $\theta_{mix}^{PA}$ are greater than the absorption coefficients $\mu_a^{mix}$ for all wavelengths in the range 680-920 nm, which was expected given the PGE of the mixtures in eq. 10. The experimental PGEs of the mixture (Fig. 2(b-h) right graph) was computed using eq 16 and were found to be almost constant with regards to the optical wavelength. Consequently, the absorption spectra multiplied by the computed median PGEs fit the photoacoustic spectra well.

For the solutions with $SO_2^{analog}$ above 90%, the PGE increased for wavelengths above 880 nm. These wavelengths are close to a minimum of absorbance for $NiSO_4$. The discrepancy between the shapes of the PA spectra and the absorption spectra for solutions of $NiSO_4$ have been noticed previously [9]. However, these discrepancies do not occur in the wavelength range of interest here (700-850 nm). Moreover, they do not affect the evaluation of the PGE of the solutions thanks to the use of a median estimator for $\widetilde{\eta_{PA}}$.

Fig. 2(e) illustrates the measurements for one of the nigrosin solutions. These solutions were selected as external references because their photoacoustic generation efficiency is expected to be equal to 1 over the entire wavelength range. We verified here that: $\widetilde{\eta_{PA}} = 100\% \pm 3\%$ (median ± MAD).

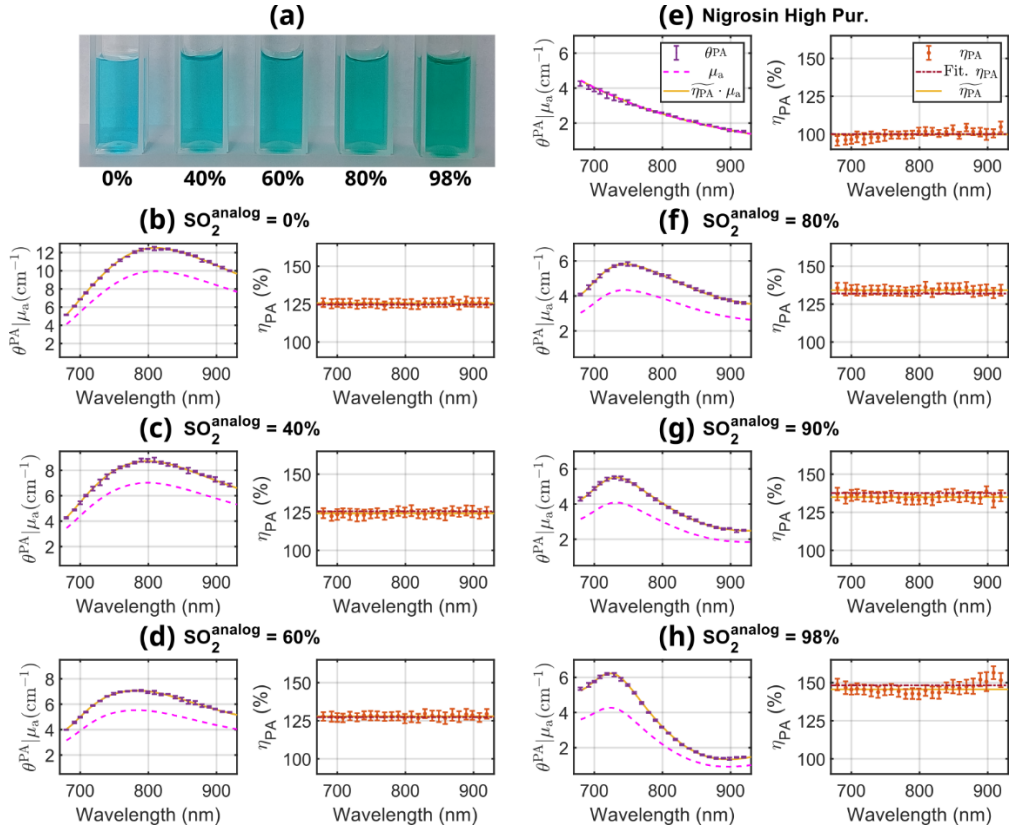

Fig. 2. (a) Photo of five mixtures modeling different $SO_2$ values and illustrating the visible color gradient. (b-h) Left sub-figure: photoacoustic coefficients $\theta^{PA}(\lambda)$ (purple error bars), absorption coefficients $\mu_a(\lambda)$ (dashed pink curve) and the absorption coefficient multiplied by the median photoacoustic efficiency: $\widetilde{\eta_{PA}} \cdot \mu_a(\lambda)$ (yellow solid line), as a function of the optical wavelength $\lambda$. All coefficients are in the same units (cm$^{-1}$) and plotted with the same vertical axis. Right sub-figure: photoacoustic generation efficiency $\eta_{PA}$ (orange error bars) as a function of the optical wavelength. The median photoacoustic generation efficiency $\widetilde{\eta_{PA}}$ was computed from the 25 measurements and displayed as a horizontal solid yellow line. The fitted $\eta_{PA}$ corresponds to the outcome of the linear regression (eq. 17 section 3.3) and is displayed as a horizontal brown dashed line.

### 3.3 Determination of the β-factors for the mixtures

Following the determination of $\widetilde{\eta_{PA}}$ for all the sample, the β-factors were determined by solving the following equation:

$$\eta_{PA}^{mix}(c_{CuSO_4}, c_{NiSO_4}) = \alpha + \beta_{CuSO_4} \cdot c_{CuSO_4} + \beta_{NiSO_4} \cdot c_{NiSO_4} \tag{17}$$

The relative uncertainty $\frac{\Delta \widetilde{\eta_{PA}}}{\widetilde{\eta_{PA}}}$ was computed and used as a weighting factor for the linear regression model (section 2.3.6). Two different methods were used to evaluate the relative uncertainty: (1) the theoretical formula for a median value of $\eta_{PA}$ knowing $\frac{\Delta \eta_{PA}(\lambda)}{\eta_{PA}(\lambda)}$ and (2) the MAD of $\eta_{PA}$. The maximum uncertainty was selected. At 25°C and for Batch #3, $\frac{\Delta \widetilde{\eta_{PA}}}{\widetilde{\eta_{PA}}}$ was found to be below 2.5% for the 10 mixtures, it reached 5% for the solutions with only NiSO$_4$ and was around 3% for the nigrosin solutions. Similar values were obtained for Batch #4 at 20°C.

Table 1 displays the results for the linear regression for all Batches and water bath temperatures. The coefficients are provided with their standard errors. The coefficient $\alpha$ is expected to be equal to one, since the Grüneisen coefficient of the mixture is equal to that of water when the concentration of sulfate salts equals zero. For Batch #1 and Batch #2, only mixtures and the calibration solution of $CuSO_4$ were included and $\alpha$ had to be set equal to 1 in the linear regression model not to bias the evaluation of the β-factors. For Batch #3 and Batch #4, nigrosin solutions were added which allowed the estimation of a coefficient $\alpha$ equal to 1. For each measurement series, the goodness of the fit was evaluated with the adjusted coefficient of determination $R^2$, the root mean squared error and the maximum absolute relative error. All these indicators confirm that the linear model is a good approximation with a maximum relative error of 2% to 4%.

All measurements in a series were consistent. However, different β-factors were determined from batch to batch at 25°C, even when the four samples with only $CuSO_4$ or $NiSO_4$ were added (Batch #3 and #4). Moreover, the β-factors were found to be similar for Batch #4 at 25°C and 20°C, while larger β-factors were determined at 20°C for Batch #3. For Batch #3 and Batch #4, $\widetilde{\eta_{PA}}$ for the nigrosin solution was consistently found to be equal to 1, and the $\widetilde{\eta_{PA}^{mix}}$ from one batch to the another were not linked by a multiplicative factor, eliminating a potential bias due to the calibration of the PA spectrometer. Sample preparation was also performed carefully, even if the preparation of new stock solutions for each batch can induce a variability. An additional variability can be induced by the fact that the absorption coefficients of the mixtures were measured at room temperature with our spectrophotometer. Petrova $et\ al$ [39] reported an increase in the absorption coefficient of solutions of $CuSO_4$ with increasing temperature, which could lead to a variability of a few percent in our case. However, across all series, β-factors were found in a 0.3 L.mol$^{-1}$ range: $\beta_{CuSO_4}$ and $\beta_{NiSO_4}$ were found within the range [0.7 L.mol$^{-1}$; 1.0 L.mol$^{-1}$] and within the range [0.55 L.mol$^{-1}$; 0.85 L.mol$^{-1}$], respectively. β-factors previously determined by our research team [9] are within these ranges, although the calibration method has been improved since earlier measurements so that estimate uncertainty has been reduced. $\beta_{CuSO_4}^{preparation\ \#1}$ as determined by Fonseca $et\ al$ [8] is also in the range that we found, but their estimate $\beta_{NiSO_4}^{preparation\ \#1}$ is not within the range we observed.

Confronted with the difficulty of precisely determining the β-factors and to obtain a robust, less precise, estimate at room temperature, we combined the 53 measured $\widetilde{\eta_{PA}^{mix}}$ to fit with equation (17). With this number of measurements from mixtures, the coefficient $\alpha$ was fitted to 1.0 without the need to add the measurements performed on nigrosin solutions. Although the quality of the fit was degraded due to variability between the measurements from different batches and temperatures, we determined that assuming $\beta_{CuSO_4}$=0.8 L.mol$^{-1}$ and $\beta_{NiSO_4}$=0.7 L.mol$^{-1}$ results in a maximum relative error of 7% in the determination of $\widetilde{\eta_{PA}^{mix}}$. This error is considered acceptable for an imaging phantom used at room temperature.

Table 1. Coefficients estimated using the linear regression model (eq. 17) fit to the $\widetilde{\eta_{PA}}$ data

| Batch # | #1 | #2 | #3 | | #4 | | All mixtures |
|---|---|---|---|---|---|---|---|
| Temperature | 25°C | 25°C | 25°C | 20°C | 25°C | 20°C | all |
| Number of samples [a] | 9 | 11 | 17 | 16 | 7 | 17 | - |
| Number of mixtures / nigrosin solutions [b] | 8 / 0 | 10 / 0 | 10 / 2 | 10 / 1 | 5 / 1 | 10 / 2 | 53 / 0 |
| Coefficients [c] | | | | | | | |
| $\alpha$ | 1 (fixed) | 1 (fixed) | 1.00 ± 0.01 | 0.99 ± 0.02 | 1.00 ± 0.02 | 1.01 ± 0.02 | **1.0 ± 0.1** |
| $\beta_{CuSO_4}$ (L.mol$^{-1}$) | 0.84 ± 0.04 | 0.70 ± 0.03 | 0.70 ± 0.05 | 0.81 ± 0.08 | 0.98 ± 0.08 | 0.95 ± 0.07 | **0.8 ± 0.1** |
| $\beta_{NiSO_4}$ (L.mol$^{-1}$) | 0.76 ± 0.03 | 0.64 ± 0.01 | 0.58 ± 0.02 | 0.77 ± 0.03 | 0.81 ± 0.03 | 0.83 ± 0.03 | **0.7 ± 0.1** |
| Model statistics [d] | | | | | | | |
| Adjusted R$^2$ | 0.975 | 0.980 | 0.978 | 0.978 | 0.989 | 0.976 | 0.7 |
| RMSE | 0.02 | 0.02 | 0.02 | 0.02 | 0.02 | 0.02 | 0.05 |
| MRE for mixtures (%) | 2 | 3 | 2 | 2 | 2 | 4 | **7** |

[a] The number of samples includes the mixtures, the nigrosin solutions and the solutions with only CuSO$_4$ or NiSO$_4$. The calibration solution of CuSO$_4$ is also included. [b] A solution prepared from the Thermo Fisher Scientific nigrosin was used when only one nigrosin solution is indicated; [c] Estimate ± standard error of the coefficient; [d] Coefficient of determination (R$^2$), Root Mean Squared Error (RMSE) and maximum relative error (MRE) of the linear regression model.

### 3.4 Summary of the parameters for the blood mimicking solutions

Table 2 presents the experimentally determined coefficients required to calculate the concentrations of the analogs in the mixture using eq. 14. For the sake of simplicity, the mass concentrations of the analogs are given for a set of representative SO$_2$ values (Table 3). From the computed concentrations, we evaluated that Batch #3 had the lowest maximum relative error for the concentrations of both analogs (about 2.5%). The concentrations for Batch #3 are also given in Table 3 and the PA measurements of Batch #3 illustrate the representative spectra mimicking of the photoacoustic coefficient of whole blood in Fig. 3.

A volume contraction is observed when dissolving the crystals of CuSO$_4$ or NiSO$_4$ in water. Therefore, mixtures, or preferably stock solutions, should be prepared in volumetric flasks from the crystals. Stock solutions at 100 g.L$^{-1}$ for CuSO$_4$ and 250 g.L$^{-1}$ for NiSO$_4$ enable the preparation of mixtures mimicking SO$_2$ levels from 0% to 100% and are well below the solubility limits (~320 g.L$^{-1}$ for CuSO$_4$ and 625 g.L$^{-1}$ for NiSO$_4$ [8]). Our recommendation is to prepare mixtures from stock solutions and to store them in sealed containers, such as polypropylene centrifuge tubes with a screw cap.

**Table 2. Determined parameters to prepare mixture solutions of the analogs**

| Analog | CuSO$_4$.5H$_2$O | NiSO$_4$.6H$_2$O |
|---|---|---|
| Molar absorption coefficient at 790 nm ε (cm$^{-1}$.L.mol$^{-1}$) | 12.05 | 1.10 |
| β factor (L.mol$^{-1}$) | 0.8 | 0.7 |
| Molar mass (g.mol$^{-1}$) [a] | 249.69 | 262.85 |
| Whole blood | SO$_2$ = 0 % | SO$_2$ = 98% |
| μ$_a$ at 760 nm (cm$^{-1}$) [b] | 8.4 | 2.7 |
| $\eta_{PA}^{wb}$ at 22°C | 1.417 | 1.417 |

[a] Given by the manufacturer. [b] From Lasers Med Sci **29**, 453–479 (2014).

**Table 3. Analog mass concentrations for whole-blood mimicking solutions at selected SO2 levels**

| | Model with coefficients of Table 2 | | Batch #3 | |
|---|---|---|---|---|
| SO$_2$ | $c_{m,CuSO_4}$ (g.L$^{-1}$) | $c_{m,NiSO_4}$ (g.L$^{-1}$) | $c_{m,CuSO_4}$ (g.L$^{-1}$) | $c_{m,NiSO_4}$ (g.L$^{-1}$) |
| 40 % | 56.7 | 39.8 | 57.8 | 40.8 |
| 60 % | 42 | 66.3 | 42.6 | 67.2 |
| 70 % | 34.1 | 83.9 | 34.6 | 84.6 |
| 80 % | 25.6 | 107.7 | 25.8 | 108.6 |
| 90 % | 15.4 | 146 | 15.4 | 146.4 |
| 98 % | 4 | 207.3 | 4 | 207.6 |

*3.5 Superimposed photoacoustic spectra of whole blood and analog mixtures*

Fig. 3 presents the superimposition of the modeled photoacoustic coefficients of whole blood for $SO_2$ levels from 40% to 98% (bottom wavelength axis) together with the spectrum of the mixtures (top wavelength axis). The wavelength axes are different due to the flip presented in section 2.2.2, but cover the same wavelength range. One can notice the spectral shape similarity between whole blood and analogs between $\lambda_1^{Hb}$ = 700 nm and $\lambda_4^{Hb}$ = 800 nm for $SO_2 \geq 80$ %. The shape similarity degrades for lower $SO_2$ values due to the absence of a sharp peak in the spectrum of $CuSO_4$ around $\lambda_3^{analog}$ =790 $nm$. The correspondence of the photoacoustic coefficients is noticeable for all modeled $SO_2$ levels at $\lambda_1^{Hb}$ = 700 nm, $\lambda_3^{Hb}$ = 760 nm (by design), and at $\lambda_5^{Hb}$ = 850 nm. The experimental data from Batch #3 at 25°C were chosen because they best match the whole blood spectrum in terms of amplitude.

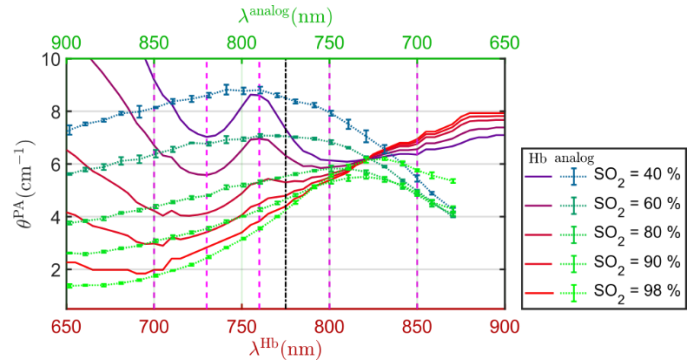

Fig. 3. Photoacoustic coefficient of whole blood (bottom x-axis and solid lines) for $SO_2$ values from 40% to 98%. The PA coefficients of whole blood were computed from eq. 8 with parameters from the literature [10,16]. Experimental spectra from Batch #3 measured at 25°C and superimposed using the top x-axis (error bars with dotted lines). The experimental data were color coded with a gradient from blue to green as $SO_2$ increases.

## 4. Discussion and conclusion

The goal of this study was to provide a practical and simple method to prepare solutions mimicking the photoacoustic properties of whole blood, that is to say the optical absorption coefficient at different oxygen saturation levels and the Grüneisen coefficient. The proposed method is inspired based on previous work [8] that suggested to use aqueous solutions of copper sulfate and nickel sulfate as analogs of oxy and deoxyhemoglobin. We went one step further by deriving an analytical formula to calculate the analog concentrations at room temperature that best matches the mixture's photoacoustic coefficient (optical absorption multiplied by the Grüneisen coefficient relative to water) to the PA coefficient of whole blood with a selected $SO_2$ level in the range from 700 nm to 850 nm. This wavelength range was identified from a literature review of the optical wavelengths most commonly used in photoacoustic imaging to probe oxygen saturation of hemoglobin. We prepared several batches of aqueous sulfate salt solutions and mixtures that enabled us to evaluate the key parameters of the molar extinction coefficient and the photoacoustic generation efficiency at room temperature. The photoacoustic generation efficiency was modeled to be a linear function of the analog concentrations, and the coefficients – β factors – were determined. The batch-to-batch variability of the β factors was mitigated to yield a satisfying approximation in the operational concentration range. Finally, the analog mass concentrations required to mimic the PA spectral response from whole blood for a broad range of oxygen saturation levels were computed to facilitate the adoption of this type of blood-mimicking solutions by the PA community. Using these solutions in imaging phantoms will enable future testing of the sensitivity and spectral performance of photoacoustic imaging systems for oxygen saturation measurement.

An originality of our approach is that it applies a mirror transformation of the optical excitation wavelengths with respect to 775 nm. This transformation maintains the boundaries of the wavelength range – 700 nm and 850 nm – while enabling a better fit of the spectral shape of $HbO_2$. However, this transformation requires a tunable laser and assumes that both the imaging system and the imaging phantom are mostly insensitive to the excitation wavelength. The background absorption of the imaging phantom is designed to be neutral, and the scattering properties are expected to be mostly homogenous for this relative wavelength range of 20% [1]. Moreover, photoacoustic imaging systems with tunable laser are expected to have similar performance at 700 nm and 850 nm. Therefore, we believe that the method will readily benefit the evaluation and optimization of blood saturation estimations with spectral photoacoustic systems. For photoacoustic systems with fixed wavelengths, the method can still be directly applied if the system wavelengths are near to 700 nm, 750 nm and 850 nm. Otherwise, a new combination of analog concentrations would have to be determined to mimic whole blood properties as a function of oxygen saturation for the necessary wavelengths and the analogy proposed by Fonseca *et al.* [8] could be tested. In all scenarii, the β factors determined in this study will enable modeling of the Gruneïsen coefficient of the mixture.

In conclusion, our results provide the practical knowledge required to use nickel sulfate and copper sulfate mixtures as analogs for oxy and deoxyhemoglobin. This method is expected to benefit the photoacoustic imaging community, especially for the development and validation of novel instruments operating at multiple wavelengths in the NIR.

**Funding.** This work was partly funded by France Life Imaging (grant ANR-11-INBS-0006) and has received financial support from 2021-2030 Cancer Control Strategy on funds administered by Inserm. L.D. acknowledges funding from the doctoral program Interfaces for Life. A.B. acknowledges funding from the Doctoral School Pierre Louis of Public health.

**Data availability.** Data underlying the results presented in this paper are not publicly available at this time but may be obtained from the authors upon reasonable request.

**References**

1. S. L. Jacques, "Optical properties of biological tissues: A review," Physics in Medicine and Biology **58**, (2013).
2. A. Taruttis and V. Ntziachristos, "Advances in real-time multispectral optoacoustic imaging and its applications," Nat Photon **9**, 219–227 (2015).
3. G. Diot, S. Metz, A. Noske, E. Liapis, B. Schroeder, S. V. Ovsepian, R. Meier, E. Rummeny, and V. Ntziachristos, "Multispectral Optoacoustic Tomography (MSOT) of Human Breast Cancer," Clinical Cancer Research **23**, 6912–6922 (2017).
4. J. Reber, M. Willershäuser, A. Karlas, K. Paul-Yuan, G. Diot, D. Franz, T. Fromme, S. V. Ovsepian, N. Bézière, E. Dubikovskaya, D. C. Karampinos, C. Holzapfel, H. Hauner, M. Klingenspor, and V. Ntziachristos, "Non-invasive Measurement of Brown Fat Metabolism Based on Optoacoustic Imaging of Hemoglobin Gradients," Cell Metabolism **27**, 689-701.e4 (2018).
5. M. Masthoff, A. Helfen, J. Claussen, W. Roll, A. Karlas, H. Becker, G. Gabriëls, J. Riess, W. Heindel, M. Schäfers, V. Ntziachristos, M. Eisenblätter, U. Gerth, and M. Wildgruber, "Multispectral optoacoustic tomography of systemic sclerosis," Journal of Biophotonics **11**, e201800155 (2018).
6. J. Park, S. Choi, F. Knieling, B. Clingman, S. Bohndiek, L. V. Wang, and C. Kim, "Clinical translation of photoacoustic imaging," Nat Rev Bioeng **3**, 193–212 (2024).
7. L. Hacker, J. Joseph, L. Lilaj, S. Manohar, A. M. Ivory, R. Tao, S. E. Bohndiek, and Members Of Ipasc, "Tutorial on phantoms for photoacoustic imaging applications," J. Biomed. Opt. **29**, (2024).
8. M. Fonseca, L. An, P. Beard, and B. Cox, "Sulfates as chromophores for multiwavelength photoacoustic imaging phantoms," Journal of Biomedical Optics **22**, 1 (2017).

9. T. Lucas, M. Sarkar, Y. Atlas, C. Linger, G. Renault, F. Gazeau, and J. Gateau, "Calibrated Photoacoustic Spectrometer Based on a Conventional Imaging System for In Vitro Characterization of Contrast Agents," Sensors **22**, 6543 (2022).
10. N. Bosschaart, G. J. Edelman, M. C. G. Aalders, T. G. Van Leeuwen, and D. J. Faber, "A literature review and novel theoretical approach on the optical properties of whole blood," Lasers Med Sci **29**, 453–479 (2014).
11. P. Kyriacou, K. Budidha, and T. Y. Abay, "Optical Techniques for Blood and Tissue Oxygenation," in *Encyclopedia of Biomedical Engineering* (Elsevier, 2019), pp. 461–472.
12. E. D. Chan, M. M. Chan, and M. M. Chan, "Pulse oximetry: Understanding its basic principles facilitates appreciation of its limitations," Respiratory Medicine **107**, 789–799 (2013).
13. M. Ascha, A. Bhattacharyya, J. A. Ramos, and A. R. Tonelli, "Pulse Oximetry and Arterial Oxygen Saturation during Cardiopulmonary Exercise Testing," Medicine & Science in Sports & Exercise **50**, 1992–1997 (2018).
14. C. Hartog and F. Bloos, "Venous oxygen saturation," Best Practice & Research Clinical Anaesthesiology **28**, 419–428 (2014).
15. D.-K. Yao, C. Zhang, K. Maslov, and L. V. Wang, "Photoacoustic measurement of the Grüneisen parameter of tissue," J. Biomed. Opt **19**, 017007 (2014).
16. Y. Villanueva-Palero, E. Hondebrink, W. Petersen, and W. Steenbergen, "Quantitative photoacoustic integrating sphere (QPAIS) platform for absorption coefficient and Grüneisen parameter measurements: Demonstration with human blood," Photoacoustics **6**, 9–15 (2017).
17. E. V. Savateeva, A. A. Karabutov, S. V. Solomatin, and A. A. Oraevsky, "Optical properties of blood at various levels of oxygenation studied by time-resolved detection of laser-induced pressure profiles," in A. A. Oraevsky, ed. (2002), pp. 63–75.
18. J. Joseph, M. R. Tomaszewski, I. Quiros-Gonzalez, J. Weber, J. Brunker, and S. E. Bohndiek, "Evaluation of Precision in Optoacoustic Tomography for Preclinical Imaging in Living Subjects," J Nucl Med **58**, 807–814 (2017).
19. X. L. Deán-Ben and D. Razansky, "Functional optoacoustic human angiography with handheld video rate three dimensional scanner," Photoacoustics **1**, 68–73 (2013).
20. J. Benavides-Lara, A. P. Siegel, M. M. Tsoukas, and K. Avanaki, "High-frequency photoacoustic and ultrasound imaging for skin evaluation: Pilot study for the assessment of a chemical burn," Journal of Biophotonics **17**, e202300460 (2024).
21. S. Zhao, J. Hartanto, R. Joseph, C.-H. Wu, Y. Zhao, and Y.-S. Chen, "Hybrid photoacoustic and fast super-resolution ultrasound imaging," Nat Commun **14**, 2191 (2023).
22. M. Sarkar, M. Pérez-Liva, G. Renault, B. Tavitian, and J. Gateau, "Motion Rejection and Spectral Unmixing for Accurate Estimation of In Vivo Oxygen Saturation Using Multispectral Optoacoustic Tomography," IEEE Trans. Ultrason., Ferroelect., Freq. Contr. **70**, 1671–1681 (2023).
23. Y. Goh, G. Balasundaram, M. Moothanchery, A. Attia, X. Li, H. Q. Lim, N. Burton, Y. Qiu, T. C. Putti, C. W. Chan, P. Iau, S. W. Tang, C. W. Q. Ng, F. J. Pool, P. Pillay, W. Chua, E. Sterling, S. T. Quek, and M. Olivo, "Multispectral Optoacoustic Tomography in Assessment of Breast Tumor Margins During Breast-Conserving Surgery: A First-in-human Case Study," Clinical Breast Cancer **18**, e1247–e1250 (2018).
24. M. Kouka, M. Waldner, and O. Guntinas-Lichius, "Multispectral optoacoustic tomography of benign parotid tumors in vivo: a prospective observational pilot study," Sci Rep **14**, 10597 (2024).
25. A. Becker, M. Masthoff, J. Claussen, S. J. Ford, W. Roll, M. Burg, P. J. Barth, W. Heindel, M. Schäfers, M. Eisenblätter, and M. Wildgruber, "Multispectral optoacoustic tomography of the human breast: characterisation of healthy tissue and malignant lesions using a hybrid ultrasound-optoacoustic approach," Eur Radiol **28**, 602–609 (2018).
26. Y. Tang, N. Wang, Z. Dong, M. Lowerison, A. Del Aguila, N. Johnston, T. Vu, C. Ma, Y. Xu, W. Yang, P. Song, and J. Yao, "Non-Invasive Deep-Brain Imaging With 3D Integrated Photoacoustic Tomography and Ultrasound Localization Microscopy (3D-PAULM)," IEEE Trans. Med. Imaging **44**, 994–1004 (2025).
27. A. Taruttis, A. C. Timmermans, P. C. Wouters, M. Kacprowicz, G. M. Van Dam, and V. Ntziachristos, "Optoacoustic Imaging of Human Vasculature: Feasibility by Using a Handheld Probe," Radiology **281**, 256–263 (2016).
28. L. M. Yamaleyeva, Y. Sun, T. Bledsoe, A. Hoke, S. B. Gurley, and K. B. Brosnihan, "Photoacoustic imaging for *in vivo* quantification of placental oxygenation in mice," The FASEB Journal **31**, 5520–5529 (2017).
29. M. Petri, I. Stoffels, J. Jose, J. Leyh, A. Schulz, J. Dissemond, D. Schadendorf, and J. Klode, "Photoacoustic imaging of real-time oxygen changes in chronic leg ulcers after topical application of a haemoglobin spray: a pilot study," J Wound Care **25**, 87–91 (2016).
30. J. R. Eisenbrey, M. Stanczak, F. Forsberg, F. A. Mendoza-Ballesteros, and A. Lyshchik, "Photoacoustic Oxygenation Quantification in Patients with Raynaud's: First-in-Human Results," Ultrasound in Medicine & Biology **44**, 2081–2088 (2018).
31. A. Buehler, M. Kacprowicz, A. Taruttis, and V. Ntziachristos, "Real-time handheld multispectral optoacoustic imaging," Opt. Lett. **38**, 1404 (2013).
32. C. J. Arthuis, A. Novell, F. Raes, J.-M. Escoffre, S. Lerondel, A. Le Pape, A. Bouakaz, and F. Perrotin, "Real-Time Monitoring of Placental Oxygenation during Maternal Hypoxia and Hyperoxygenation Using Photoacoustic Imaging," PLoS ONE **12**, e0169850 (2017).

33. C. Lutzweiler, R. Meier, E. Rummeny, V. Ntziachristos, and D. Razansky, "Real-time optoacoustic tomography of indocyanine green perfusion and oxygenation parameters in human finger vasculature," Opt. Lett. **39**, 4061 (2014).
34. D. J. Lawrence, M. E. Escott, L. Myers, S. Intapad, S. H. Lindsey, and C. L. Bayer, "Spectral photoacoustic imaging to estimate in vivo placental oxygenation during preeclampsia," Sci Rep **9**, 558 (2019).
35. B. F. Combes, S. K. Kalva, P.-L. Benveniste, A. Tournant, M. H. Law, J. Newton, M. Krüger, R. Z. Weber, I. Dias, D. Noain, X. L. Dean-Ben, U. Konietzko, C. R. Baumann, P.-G. Gillberg, C. Hock, R. M. Nitsch, J. Cohen-Adad, D. Razansky, and R. Ni, "Spiral volumetric optoacoustic tomography of reduced oxygen saturation in the spinal cord of M83 mouse model of Parkinson's disease," Eur J Nucl Med Mol Imaging **52**, 427–443 (2025).
36. M. Toi, Y. Asao, Y. Matsumoto, H. Sekiguchi, A. Yoshikawa, M. Takada, M. Kataoka, T. Endo, N. Kawaguchi-Sakita, M. Kawashima, E. Fakhrejahani, S. Kanao, I. Yamaga, Y. Nakayama, M. Tokiwa, M. Torii, T. Yagi, T. Sakurai, K. Togashi, and T. Shiina, "Visualization of tumor-related blood vessels in human breast by photoacoustic imaging system with a hemispherical detector array," Sci Rep **7**, 41970 (2017).
37. X. L. Deán-Ben, T. F. Fehm, M. Gostic, and D. Razansky, "Volumetric hand-held optoacoustic angiography as a tool for real-time screening of dense breast," J. Biophoton **9**, 253–259 (2016).
38. T. Lucas, C. Linger, T. Naillon, M. Hashemkhani, L. Abiven, B. Viana, C. Chaneac, G. Laurent, R. Bazzi, S. Roux, S. Becharef, G. Avveduto, F. Gazeau, and J. Gateau, "Quantitative, precise and multi-wavelength evaluation of the light-to-heat conversion efficiency for nanoparticular photothermal agents with calibrated photoacoustic spectroscopy," Nanoscale **15**, 17085–17096 (2023).
39. E. Petrova, S. Ermilov, R. Su, V. Nadvoretskiy, A. Conjusteau, and A. Oraevsky, "Using optoacoustic imaging for measuring the temperature dependence of Grüneisen parameter in optically absorbing solutions," Optics Express **21**, 25077 (2013).